\documentclass[journal=jacsat,manuscript=communication,articletitle=true]{achemso}
\usepackage[version=3]{mhchem} 
\usepackage{booktabs}
\usepackage{graphicx}
\usepackage{siunitx}
\usepackage{amsmath}
\usepackage{amssymb}
\usepackage{amsfonts}
\usepackage{epsfig}
\usepackage{ulem}

\SectionsOn
\SectionNumbersOn

\author{Jo\~ao Paulo Almeida de Mendon\c ca}
\email{joao-paulo.almeida-de-mendonca@grenoble-inp.fr}
\affiliation[SIMaP]
{Université Grenoble Alpes, CNRS, Grenoble INP, SIMaP, 38000 Grenoble, France}

\author{Lorenzo Antonio Mariano}
\affiliation[SIMaP]
{Université Grenoble Alpes, CNRS, Grenoble INP, SIMaP, 38000 Grenoble, France}

\author{Emilie Devijver}
\affiliation[LIG]
{Université Grenoble Alpes, CNRS, Grenoble INP, LIG, 38000 Grenoble, France}

\author{Noel Jakse}
\affiliation[SIMaP]
{Université Grenoble Alpes, CNRS, Grenoble INP, SIMaP, 38000 Grenoble, France}

\author{Roberta Poloni}
\email{roberta.poloni@grenoble-inp.fr}
\affiliation[SIMaP]
{Université Grenoble Alpes, CNRS, Grenoble INP, SIMaP, 38000 Grenoble, France}

\title[Density Functional for Adiabatic Energy Differences]{An Artificial Neural Network-based Density  Functional Approach for Adiabatic Energy Differences in Transition Metal Complexes}

\begin{document}

\begin{tocentry}
\end{tocentry}

\begin{abstract}
During the past decades, approximate Kohn-Sham density-functional theory schemes garnered many successes in computational chemistry and physics; yet the performance in the prediction of spin state energetics is often unsatisfactory.  
 By means of a machine-learning approach, an enhanced exchange and correlation functional is developed to describe adiabatic energy differences in transition metal complexes.
The functional is based on the computationally efficient revision of the regularized strongly constrained and appropriately normed (R2SCAN) functional and improved by an artificial neural-network-correction trained over a small dataset of electronic densities, atomization energies and/or spin state energetics. 
The training process, performed using a bio-inspired non gradient-based approach adapted for this work from Particle Swarm Optimization, is discussed in detail. The meta-GGA functional is finally shown to outperform most known density functionals in the prediction of adiabatic energy differences for both the validation test and the generality test.
\end{abstract}

\section{Introduction}
Applied density-functional theory (DFT) has revolutionized the field of materials science by providing a powerful tool for predicting the electronic and structural properties of materials, enabling their design and discovery for a wide range of applications\cite{RevModPhys}. Despite its many successes, DFT still faces several challenges that limit its predictive power and applicability \cite{Weitao,Pol2018}. As an example, one can metion the calculation of spin-state energetics in transition metal complexes\cite{wilbraham2017multiconfiguration,domingo2010spin,radon2019benchmarking,swart2008accurate,phung2018toward,Rei2023}.  This has strong implications in the assistance and guidance alongside experimental efforts towards, for example, the design of novel spin crossover (SCO) complexes. 
 The vast majority of SCO materials, which are of great interest for applications such as molecular spintronics, molecular electronics,  and sensors\cite{molnar2019molecular,kumar2017emerging,C9SC02522G,D0DT01533D} are octahedrally coordinated molecular complexes. 
 For these complexes, the thermodynamics of the spin transition (\textit{i.e.}, the transition temperature T$_{1/2}$) between the low-spin (LS) and high-spin (HS) states is related to the adiabatic energy difference, $\Delta E$\textsubscript{H-L}=$E$\textsubscript{HS}-$E$\textsubscript{LS}, \textit{i.e.}, the energy difference between the two spin states computed at their corresponding geometry. 
 
 The computational chemistry community has devoted a substantial effort in understanding the performance of approximate Khon-Sham density-functional theory (KS-DFT) schemes  \cite{swart2004validation,pierloot2006relative,swart2008accurate,droghetti2012assessment,song2018benchmarks}. 
 Large deviations in the values of the spin splitting energies are found among different families of exchange and correlation functionals. 
 Semilocal functionals, such as the generalized gradient approximation (GGA), for example, tend to overstabilize LS, \cite{swart2004validation,Fouqueau2005,radon2014revisiting,mortensen2015spin,ioannidis2015towards,kulik2015perspective} while Hartree–Fock (HF) overstabilizes HS \cite{Rehier2012}. In this context, hybrid functionals may be viewed as a possibility to alleviate the problem \cite{Rehier2012,Prokopiou2018,song2018benchmarks}.
DFT+U was also studied in this respect\cite{kulikmarzari2006,CococcMarzari2007,kulik2010systematic,Vela2020,Uff2022}. Recently, it was shown that the Hubbard $U$ correction in DFT+U may yield large errors in the spin-state energy difference of octahedral Fe-II complexes owing to the bias toward high-spin states imposed by the Hubbard term in the total energy\cite{mariano2020}. Later, a non-self consistent density-corrected scheme adopting the Hubbard $U$ density was  shown to yield a good description of adiabatic energy differences for Fe(II)-based complexes\cite{mariano2021improved}.



Recently, several studies have shown how density functional theory can benefit from machine learning (ML) techniques~\cite{nagai2022machine,li2021kohn,kirkpatrick2021pushing,king2021machine, dick2021highly,nagai2020completing,chen2020deepks,dick2020machine,nagai2018neural, brockherde2017bypassing,li2016understanding,snyder2012finding}. Pioneer works like the one from Snyder \textit{et al. } proved that is possible to learn the kinetic energy of 1D fermionic systems\cite{snyder2012finding}. Brockherde \textit{et al.} developed a ML scheme to learn the density \textit{via} the external potential/density Hohenberg--Kohn map, so that self consistency can be bypassed~\cite{brockherde2017bypassing}. Learning the exchange and correlation functional itself, as demonstrated by the pioneer work by Nagai \textit{et al.}\cite{nagai2020completing}, has also been studied \cite{nagai2020completing, chen2020deepks,li2021kohn,dick2021highly,kirkpatrick2021pushing,nagai2022machine}. In this class of methods, the exchange and correlation is replaced, adjusted, or corrected by an artificial neural network that receives as input functions of the electronic density. 

These machine learning applications to KS-DFT use the supervised  approach philosophy where the functional "learns itself" from a given high quality set of data~\cite{kalita2021learning, li2016understanding}.
 An important aspect worth to be mentioned here is   that the knowledge of the physical constraints that the exact exchange and correlation functional must fulfill can be exploited to create a strongly transferable functional.
 In this spirit, Nagai, Akashi, and Sugino~\cite{nagai2022machine} trained their ANN on the atomization energies and densities from single point calculations of three molecules, and showed that their resulting metal-GGA functional performed well in describing atomization energies of molecules not seen in the training, and also on computing equilibrium lattice constants for solids.

 The aim of the present work is to extend upon the latter approach and train an exchange and correlation functional for a good description of electronic densities and adiabatic energy differences of transition metal complexes, as well as atomization energies and densities of simpler molecules.  An important novelty of this work, is that the training process is explicitly shown and discussed in detail, thus allowing the reader to understand the optimization process and its limitations.
The ML functional is trained using a bio-inspired non gradient-based approach adapted from Particle Swarm Optimization (PSO)\cite{kennedy1995particle}. 
This class of optimization algorithms has shown success in handling intricate nonlinear loss functions\cite{tian2022modern} owing to the collective intelligence ingredient that improves the efficiency of the optimization process. 
These results show that by training a correction over R2SCAN using three light molecules and three diatomic (metal-nonmetal) transition molecules, the prediction of adiabatic energy differences are improved compared to the state-of-the-art in approximate KS-DFT at no expenses for the performance on atomization energies. 
Furthermore, the analysis of the non-gradient based training shows that the $\Delta E_{H-L}$ term dominates the optimization and that the electronic density plays an auxiliary role by accelerating the convergence of the loss function.  
Compared to the known density functional approaches tested here for comparison, the exchange and correlation functional optimized in this work exhibits the best performance for both the validation and the generality test.


\section{Methodology}

\subsection{Density Functional Theory and Neural Network Functionals}

In the KS-DFT framework, the total energy, $E_\textup{tot}$, and the electronic density, $\rho(\textbf{r})$, can be computed by solving the KS equations for non-interacting fermions\cite{kohn1965self}
\begin{equation}
    \label{eq:ks}
    \left(-\frac{\hbar}{2m}\nabla^2+v_\textup{eff}\right)\phi_i=\varepsilon_i\phi_i.
\end{equation}
In Eq.~\ref{eq:ks}, the effective potential reads
\begin{equation}
    \label{eq:veff}
    v_\textup{eff}(\textbf{r})=v_\textup{ext}(\textbf{r})
    +\int \frac{\rho(\textbf{r})}{|\textbf{r}-\textbf{r}'|}d\textbf{r}'
    +\frac{\delta E_\textup{xc}[\rho(\textbf{r})]}{\delta\rho(\textbf{r})},
\end{equation}
where $\rho(\textbf{r}) = \sum \phi_i\phi_i^{*}$, $v_\textup{ext}(\textbf{r})$ is an external potential and $E_\textup{xc}$ is the functional of exchange and correlation. The exact form of $E_\textup{xc}$ is unknown and therefore must be approximated. By solving the eigen-equation presented in Eq.{~}\ref{eq:ks}, one obtains the eigenvalues, $\varepsilon_i$, and eigenfunctions, $\phi_i$, that correspond to the Kohn-Sham energies and orbitals. Because $\rho(\textbf{r})$ is needed to build $v_\textup{eff}$ and therefore compute $E_\textup{tot}$, the set of equations above is solved self-consistently, \textit{i.e.}, $\rho(\textbf{r})$ is recomputed at iterations until convergence on the total energy and charge density is achieved. The dependency of the effective potential $v_\textup{eff}$ on $\rho$ is at the core of the mean-field theory of KS-DFT since it should mimic the electronic correlation of the many electrons system by adopting a non-interacting single particle scheme.

After solving the KS equations, the total energy can be written as
\begin{equation}
\label{eq:etot}
\begin{split}
    E=&\sum \varepsilon_i -
    \frac{e^2}{2}\iint \frac{\rho(\textbf{r})\rho(\textbf{r}')}{|\textbf{r}-\textbf{r}'|} d\textbf{r}'d\textbf{r}
    + E_\textup{xc}[\rho(\textbf{r})] \\
    &- \int \frac{\delta E_\textup{xc}[\rho(\textbf{r})]}{\delta \rho(\textbf{r})}\rho(\textbf{r}) d\textbf{r}.
\end{split}
\end{equation} 

\begin{figure*}[ht]
    \centering
    \includegraphics[width=0.9\linewidth]{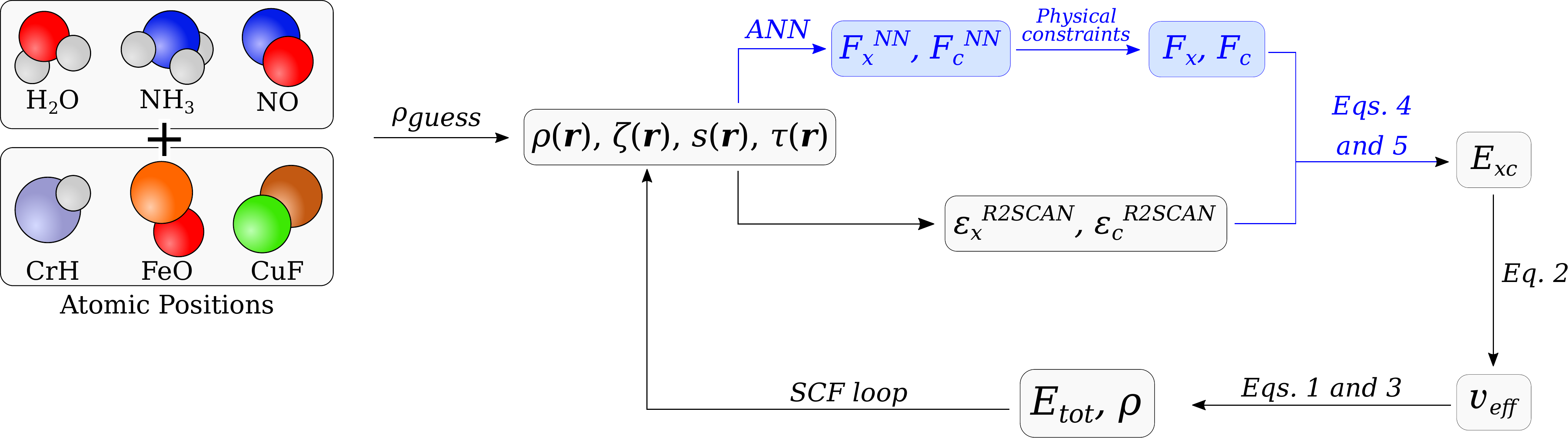}
    \caption{Schematic representation of a single point calculation performed in the present approach. The step where the ANN correction is explicitly applied over R2SCAN is highlighted in blue.}
    \label{fig:dft}
\end{figure*}

The functional of exchange and correlation can be written as  
\begin{equation}
\label{eq:Exc}
E_\textup{xc}=\int \varepsilon_\textup{xc}[\rho(\textbf{r})] \rho(\textbf{r}) d\textbf{r},
\end{equation}
where $\varepsilon_\textup{xc}$ is  the exchange and correlation energy density. The idea developed in this work, summarized in Figure{~}\ref{fig:dft}, is to obtain a $\varepsilon_\textup{xc}$ as a correction over R2SCAN functional,\cite{furness2020accurate} such that
\begin{equation}
\label{eq:exc}
\varepsilon_\textup{xc}=F_{x}\varepsilon_\textup{x}^\textup{R2SCAN}+F_\textup{c}\varepsilon_\textup{c}^\textup{R2SCAN}.
\end{equation}

Recently,  Nagai and co-workers\cite{nagai2020completing} developed a new functional, defined as a correction over the strongly constrained and appropriately normed (SCAN)\cite{sun2015strongly}, trained on energies and densities of a small subset (three molecules only made up by first and second row elements) of the G2 dataset\cite{curtiss1997assessment}.
The result showed an improvement over SCAN in the prediction of ionization potentials and atomization energies on the complete G2 dataset. Yet, previous reports on the literature pointed out some major numerical instabilities in some meta-GGA functionals \cite{furness2019enhancing} including SCAN itself, for electron densities rapidly changing in space. These manifest as sharp features of the energy derivatives and make the 
functional unstable over grid changes thus resulting in bad gradients estimation during the  self-consistent field  process~\cite{yang2016more,sitkiewicz2022reliable}.
To solve this problem, Bartok and co-workers developed a regularized version of SCAN (rSCAN) which  exhibit improved smoothness properties and improves SCAN's numerical performance~\cite{bartok2019regularized}. However, this improvement came at the expense of some of the physical limits originally obeyed by SCAN. R2SCAN\cite{furness2020accurate} follows this implementation path, recovering most of the physical limits that SCAN originally had. The $17$ exact conditions applicable to meta-GGA (listed in the SI from \citet{sun2015strongly}) were imposed in the original definition of SCAN and most of these conditions are present on R2SCAN. Only the fourth-order gradient expansion condition, lost in the rSCAN implementation, is not recovered.\cite{furness2020accurate}

In this work, in order to address the case of transition metal complexes, where rapid fluctuations of the electronic density are expected, a new ANN-based functional defined as a correction over R2SCAN is proposed. 
This choice allows us to circumvent the instabilities of SCAN that manifest themselves in large changes of the energy upon tiny changes of the density,  as shown in Figure S1 for the \ce{FeO} molecule. Such instability (observed for FeO and not for H$_2$O) is found when training a correction over SCAN.
Finally, we impose that
$\varepsilon_\textup{xc}$, defined in eq.~\ref{eq:exc}, satisfies the physical limits imposed in R2SCAN using Lagrange multipliers, as done by Nagai and co-workers in their more recent work.\cite{nagai2022machine}

\begin{figure}[ht]
    \centering
    \includegraphics[width=0.9\linewidth]{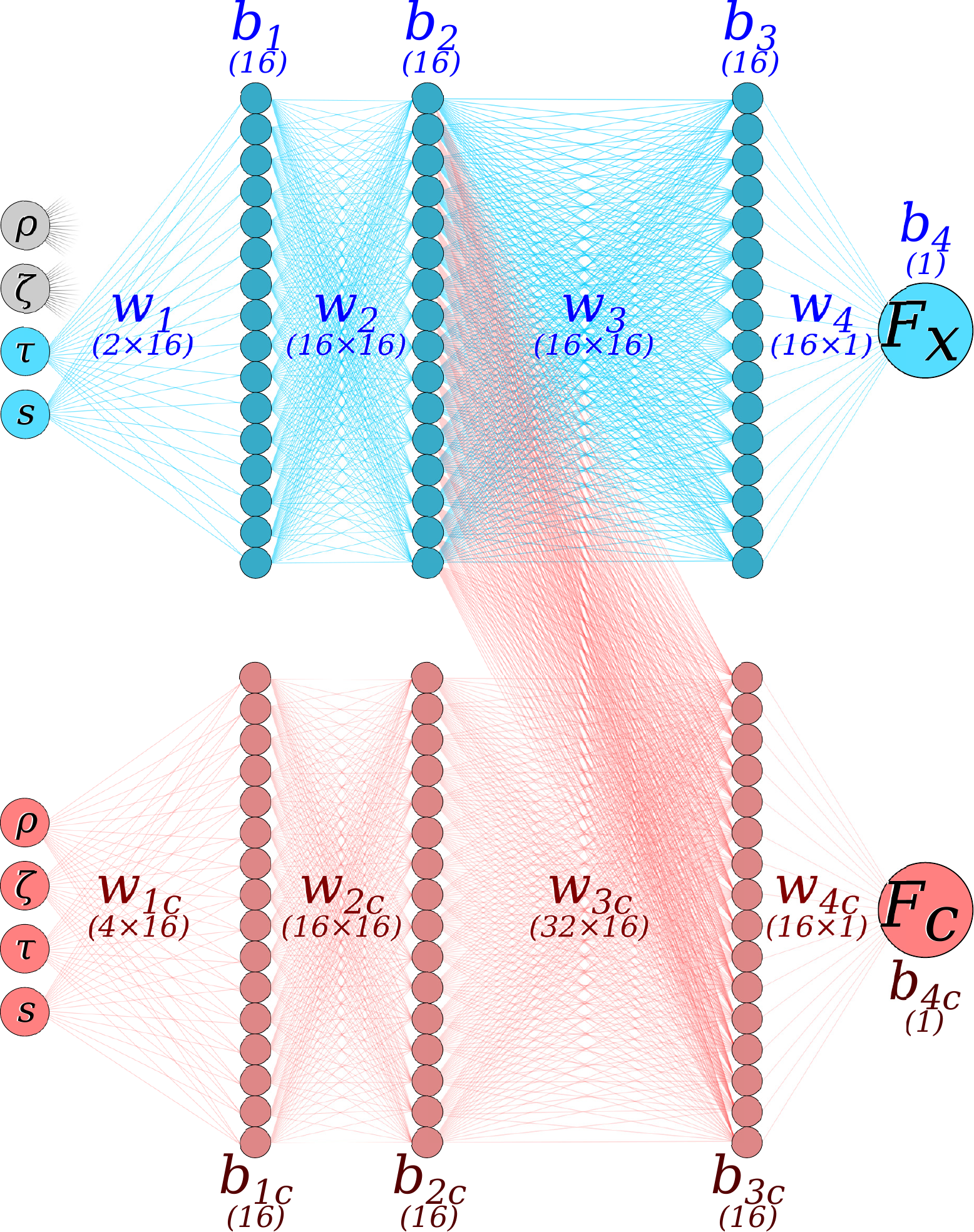}
    \caption{Architecture of the artificial neural network used in the present work. The $\textsl{\textrm{w}}$'s and $b$'s represent the weights and bias used on the feed-forward process and the products in parenthesis express the number of parameters in each of those elements.}
    \label{fig:arch}
\end{figure}

Figure{~}\ref{fig:arch} shows a diagrammatic representation of the feed-forward neural network adopted here. The architecture reflects the choice to adopt a small number of fitted parameters.
Recent efforts on ML functionals are moving in the direction of complex and deeper neural networks which  results in a large amount of parameters to be fitted. For example, Nagai \textit{et al.} use an architecture with hidden layers composed of $100$ neurons each, resulting in a the total number of parameters surpassing $10^4$~\cite{nagai2022machine}. The recent DM21 functional from DeepMind also report a total of roughly $4\times10^5$ parameters~\cite{kirkpatrick2021pushing}. Here, the choice was made to move in the opposite direction, adopting a smaller neural network with precisely $1506$ parameters, which is more efficient to train. 

The local and semilocal functions that are provided to the input neurons of the ANN are consistent with a meta-GGA functional. Evaluated for each point in the integration grid, the inputs are defined as
\begin{equation}
\label{eq:input}
\begin{split}
\rho:=&\sum_i^\textup{occ}{|\phi_i|}^2,\\
\zeta:=&\frac{\rho_{\uparrow}-\rho_{\downarrow}}{\rho},\\
s:=&\frac{|\nabla\rho|}{2\sqrt[3]{3\pi\rho^4}},\\
\tau:=&\frac{1}{2}\sum_i^\textup{occ}|\nabla\phi_i|^2.
\end{split}
\end{equation}
These quantities correspond to the electronic density, the spin density, the scaled density gradient, and the kinetic-energy density associated to the electronic state of the system, respectively.  Pre-processing of these functions is performed with a $\textup{htan}$ based filter before they enter the ANN \cite{nagai2020completing,nagai2022machine}.
This is done \textit{via} the transformations $\overline{n_s}=\textup{htan}\left( (\tfrac{\rho}{\textup{e}})^{\tfrac{1}{3}}\right)$, $\overline{\zeta}=\textup{htan}\left( \tfrac{1}{2}\{ (1+\zeta)^{\tfrac{4}{3}}-(1+\zeta)^{\tfrac{4}{3}} \}\right)$, $\overline{s}=\textup{htan}\left(s\right)$, and $\overline{\tau}=\textup{htan}\left(\tfrac{\tau -\tau_{\textup{unif}}}{\tau_{\textup{unif}}}\right)$; where $\tau_{\textup{unif}}=\tfrac{3}{10}(3\pi^2)^{\tfrac{2}{3}}\left(\tfrac{\rho}{\textup{e}}\right)^{\tfrac{5}{3}}$.

The corrections $F_\textup{x}$ and $F_\textup{c}$ are computed separately so that different asymptotic limits are applied to each of them~\cite{sun2015strongly,nagai2022machine}. For the correct uniform coordinate density-scaling condition\cite{levy1985hellmann} and the exact spin scaling relation\cite{oliver1979spin} to be satisfied simultaneously, the exchange energy should not depend explicitly in $\rho$ and $\zeta$, so this values are not included as inputs in the calculation of $F_x$. The weights $\textsl{\textrm{w}}_{3c}$ are set as a $32\times16$ matrix, using the second hidden layer from both the exchange and the correlation networks to compute the third hidden layer of the correlation network. This improves the training process\cite{nagai2022machine} and is only possible in this direction, since the inputs for $F_x$ are a part of the inputs for $F_c$. The activation function used here is a version of the softplus function modulated so it has value and derivative as \num{1} at the origin. In this way, if all parameters on the NN where set to zero, the functional obtained would be exactly R2SCAN.

\subsection{Training of the Neural Network Functional}

\subsubsection{Loss Function}
The loss function is the sum of errors of adiabatic energy differences, atomization energies, and electron densities
\begin{equation}
\begin{split}
\Delta _\text{err}=& 
\sum_i^{\textit{G2}} 
\left[ \frac{1}{E_0}\big | AE_i^{\text{REF.}}-AE_i^{\text{DFT}} \big| 
+ \right.\\ 
& \quad  \left.
+\frac{1}{N_i}\int \big |\rho_i^{\text{REF.}}-\rho_i^{\text{DFT}}\big |dV 
\right] + \\
& \sum_j^\textit{TMC}\left[ \frac{1}{E_0} \big|\Delta E_{\text{H-L},j}^{\textup{REF.}}-\Delta E_{\text{H-L}, j}^{\textup{DFT}}\big| + \right. \\
& \quad   \left.\frac{1}{N_j}\int\left(\big|\rho_{\text{H},j}^{\textup{REF.}}-\rho_{\text{H},j}^{DFT}\big| +\big|\rho_{\text{L},j}^{\textup{REF.}}-\rho_{\text{L},j}^{\textup{DFT}}\big|\right )dV\right],
\end{split}    
\label{eq:lossfunction}
\end{equation}
where the first summation is performed over three small molecules (\ce{NO}, \ce{H2O}, \ce{NH3}) taken from the original G2 set with experimental reference values of atomization energies, $AE^{\textup{REF.}}$, extracted from Curtiss \textit{et al.}~\cite{curtiss1997assessment}.  To avoid confusion, from now on italic \textit{G2} refers to this set of three molecules, while the (non-italic) G2 refers to the complete set by Curtiss \textit{et al.}~\cite{curtiss1997assessment}.

The second summation is performed over three diatomic transition metal complexes (\textit{TMC}) with reference values for adiabatic energy difference, $\Delta E_{H-L}^{\textup{REF.}}$, taken from experiments and close to the 
CCSD(T) values within \SI{0.02}{\electronvolt}\cite{kolik2010systematic}. These molecules are \ce{FeO}, \ce{CuF}, \ce{CrH}.
The error in the density for both the \textit{G2} set and the \textit{TMC} is computed using the $L^2$ metric,
by taking the CCSD(T) reference for the  electronic density. Finally, $E_0 = 1$ Hartree, and $N_j$ is the number of electrons in the system $j$, both used to make the loss function dimensionless.  

In order to evaluate the effects of the database, six different training sets were considered: \textit{G2} ([\ce{NO} + \ce{H2O} + \ce{NH3}]), \textit{G2+FeO}, \textit{G2+CuF}, \textit{G2+CrH}, \textit{TMC} ([\ce{FeO} + \ce{CuF} + \ce{CrH}]), and \textit{ALL} (\textit{G2} + \textit{TMC}). 
The terms of the loss function only apply when at least one of the related molecules is present in the training set. 
 The Spearman correlation matrices and the principal component analysis of the loss function terms during the optimization are shown in the SI (Figures S2-S19).

The inclusion of the first term in eq.~\ref{eq:lossfunction} was shown to yield good results for energetics of light element-molecules \cite{dick2021highly,nagai2020completing}. Besides, it was shown that training sets with different types of data per sample (\textit{i.e.}, sparse training data) can significantly improve the performance of a machine learned-density functional\cite{kasin2021learning}.

The experimental values of $\Delta E_{H-L}$ are \SI{1.81}{\electronvolt} for \ce{CuF} (spin change: $S=1 \to 3$),\cite{dufour1982electronic,brazier1983nuclear} \SI{0.14}{\electronvolt} for \ce{FeO} ($S=5 \to 7$),\cite{drechsler1997mass} and \SI{-1.39}{\electronvolt} for \ce{CrH} ($S=4 \to 6$) \cite{stevensmiller1987laser,ram1993fourier}. The choice of these specific molecules is done after considering four points: (i) the values of $\Delta E_{H-L}$ span a large range from negative to positive values, (ii) the diversity in terms of composition, (iii) the adiabatic energy difference involves different spin states in each case, and (iv) the configuration interaction weight of the dominant electronic configuration computed using CASSCF is different in the three cases pointing to different degrees of static correlations ($C_0^2=0.810$, $0.783$, and $0.938$; for \ce{CuF}, \ce{FeO}, and \ce{CrH} respectively)\cite{multiref2012}. We do not claim here that our approach is capable to treat systems with multiconfigurational character. Rather, that the ANN-DFT yields a solution similar to the coupled cluster one, which gives an excellent agreement with experiment  for these molecules~\cite{kolik2010systematic}.

\subsubsection{Optimization Algorithm}

\begin{figure*}[ht]
    \centering
    \includegraphics[width=0.8\textwidth]{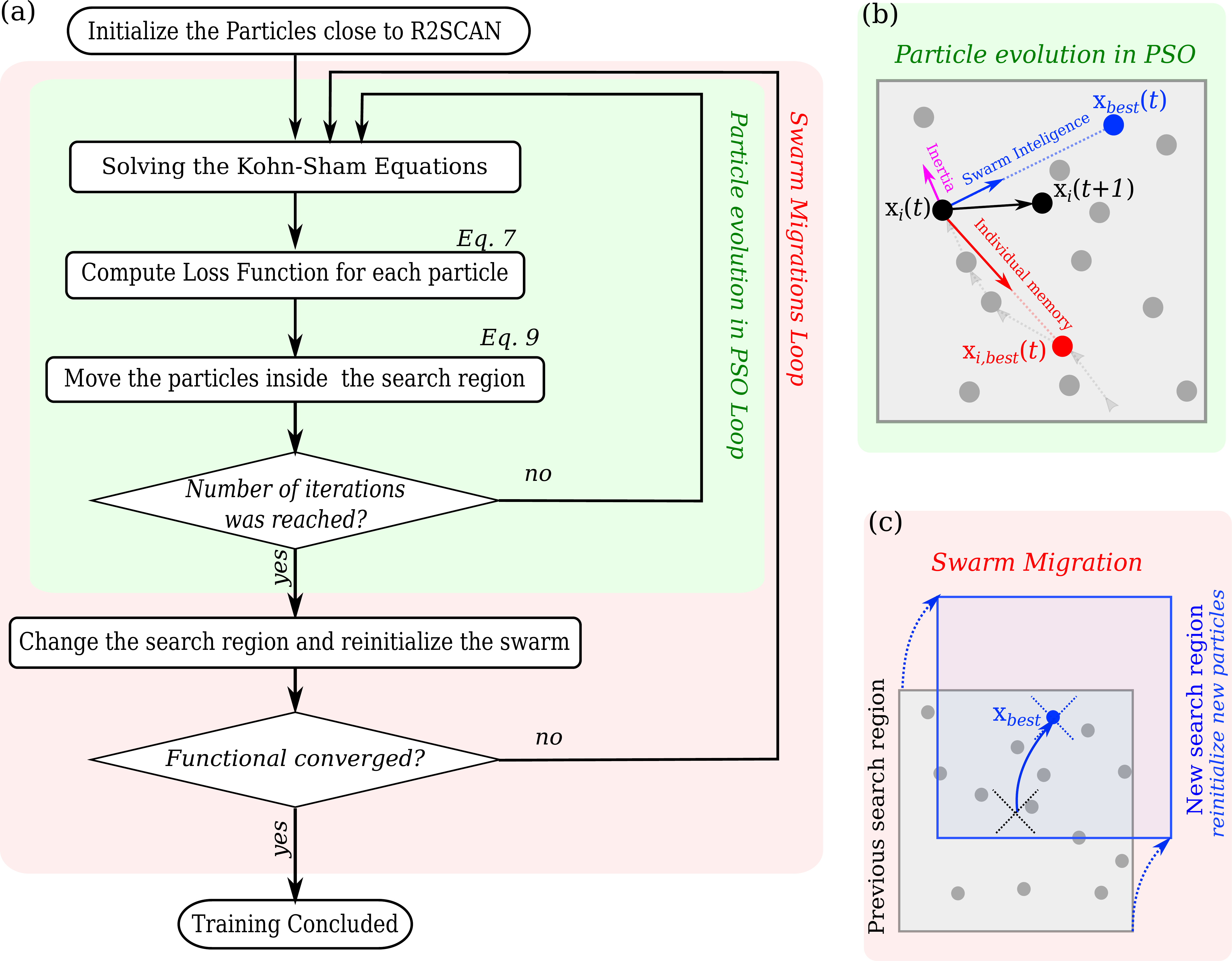}
    \caption{(a) Flowchart for the functional training approach as described in the text, including the PSO loop (green) and the migrations scheme (red). (b) Schematic representation of a PSO step. Magenta, red and blue vectors are associated with the right-hand side terms of equation{~}\ref{eq:v_pso}, respectively. (c) Migration of the search region: all particles (grey and blue circles) are removed from the previous box (grey square) and reinitialized randomly inside a new one (blue square).}
    \label{fig:flow}
\end{figure*}

 Since the equilibrium energies and densities are computed after converging the SCF process with a given set of parameters, the loss function depends on quantities that are related in a non-trivial way to the parameters of the ANN-functional. Thus, optimizing the parameters using a gradient based approach, although possible, as shown by Dick et al. and coworker\cite{dick2021highly}, may be impractical in general. Recent studies have shown that non-gradient based optimization techniques can outperform gradient-based approaches for complicated loss functions \cite{tian2022modern}. Here, an adaptation of Particle Swarm Optimization (PSO)\cite{kennedy1995particle} is used as the training algorithm. PSO is built over a bio-inspired meta-heuristic: solutions (points on the parameters space of our neural network) will be treated as particles in a swarm (a collection of solutions) that forage for resources (that search for the minima of the loss function). The application of the PSO algorithm to the problem of functional optimization is shown in the 
 flowchart of Figure{~}\ref{fig:flow}(a).
In the initial step, a search region in the parameters space is defined, and a swarm of particles with positions $\mathbf{x}_i(t=0)$ is randomly generated. These are initialized with velocities $\mathbf{v}_i(t=0)=0$ in this space. At each step $t$ in the PSO evolution, all the particles are submitted to a collective loss function evaluation. 
From this, the best solution ever visited by the swarm $\mathbf{x}_\textup{best}(t)$ and the best solution ever visited by the each particle $\mathbf{x}_{i,\textup{best}}(t)$ are saved. The velocities and positions on the swarm at {\sl{t}}+1 are then updated as follows:

\begin{equation}
\label{eq:v_pso}
\begin{split}
    \mathbf{v}_i(t+1) = & \quad\omega \mathbf{v}_i(t) \\
    &+ c_1 \textup{Rand}{(0,1]} ( \mathbf{x}_{i,\textup{best}}(t) - \mathbf{x}_i(t) )\\
    &+ c_2 \textup{Rand}{(0,1]} ( \mathbf{x}_{\textup{best}}(t) - \mathbf{x}_i(t)),
\end{split}
\end{equation}
\begin{equation}
    \mathbf{x}_i(t+1)=\mathbf{x}_i(t)+\mathbf{v}_i(t+1),
\end{equation}
where $\textup{Rand}{(0,1]}$ are random numbers independently generated in the ${(0,1]}$ interval and $\omega$, $c_1$, and $c_2$ are constants that dictate the behavior of the particles on the swarm. $\omega$ mimics the inertia of a flying particle and is associated with the pink component of the velocity in Figure{~}\ref{fig:flow}(b). $c_1$ gives to the particle a velocity component that points toward the best solution known by the particle itself and it is the red component in Figure{~}\ref{fig:flow}(b). $c_2$ gives a velocity component that points toward the best solution known by the swarm, which implements the concept of collective intelligence and it is shown by the blue component in Figure{~}\ref{fig:flow}(b).

The standard PSO approach\cite{kennedy1995particle} is modified here by changing the search region between PSO runs as described below. This solves the observed instability when trying to evaluate the loss function in points that are too far from solutions that we know already. 
At the beginning of the optimization process, the particles are initialized inside a box of size $s$ around the origin ($\varepsilon_{x/c}=F_{x/c}(0)\times\varepsilon_{x/c}^\textup{R2SCAN}=\varepsilon_{x/c}^\textup{R2SCAN}$). Once a given number of PSO steps is done, the searching box is moved so that its center corresponds to the best solution found until then as shown in Figure{~}\ref{fig:flow}(c). The particles are redistributed and the PSO search restarts. The PSO steps are performed using $c_1=0.1$, $c_2=0.2$, and $\omega=0.1$. The search box in each migration is defined using $s=0.05$ (with periodic boundary conditions) and $10$ PSO steps are taken in each migration. A total of $32$ particles are used in each step and a total of $50$ migrations is performed for every run. 

\subsection{Technical Details}

Our functional is implemented using an adaptation of the \texttt{xc\_pcNN} import provided by Nagai and co-workers:\cite{pcNN_mol} the neural network architecture is changed to the one proposed here and the physical constraints and corrections were adapted to the case of R2SCAN (imported from \texttt{libxc}\cite{lehtola2018recent}). 

The training code is an in-house application developed in \textsc{Python3}, combining the functionalities of \texttt{pyscf}\cite{sun2020recent} and \texttt{ase}\cite{larsen2017atomic} to compute the DFT SCFs and \texttt{pyswarm}\cite{miranda2018pyswarm} to perform the optimization inside each migration during the training process. For each particle (each parametrization of the ANN-DFT) the SCF is performed in order to compute the loss function. For this we use the 6-311++G(3df,3pd) for \textit{G2} and the ccpcvtz basis set for the transition metal complexes. The number of iterations within the SCF is set to 50 and the convergence criteria on the energy change is set to $10^{-6}E_\textup{h}$. 
The calculations that failed to converge within this criterion were not considered during the optimization process. 

The reference densities are obtained \textit{via} orbital optimized-CCSD(T), as implemented in ORCA 5.0\cite{neese2020orca}. All the validations on the transition metal complexes are performed using \texttt{pyscf} and \texttt{ccpcvtz}. 
The validation on the whole G2 set of molecules is (shown in section S4 the SI) use geometries form \texttt{ase.collections.g2} and the 6-311++G(3df,3pd) basis set.

\section{Results}

\subsection{Trainings}

\begin{figure*}[ht]
    \centering
    \includegraphics[width=0.95\linewidth]{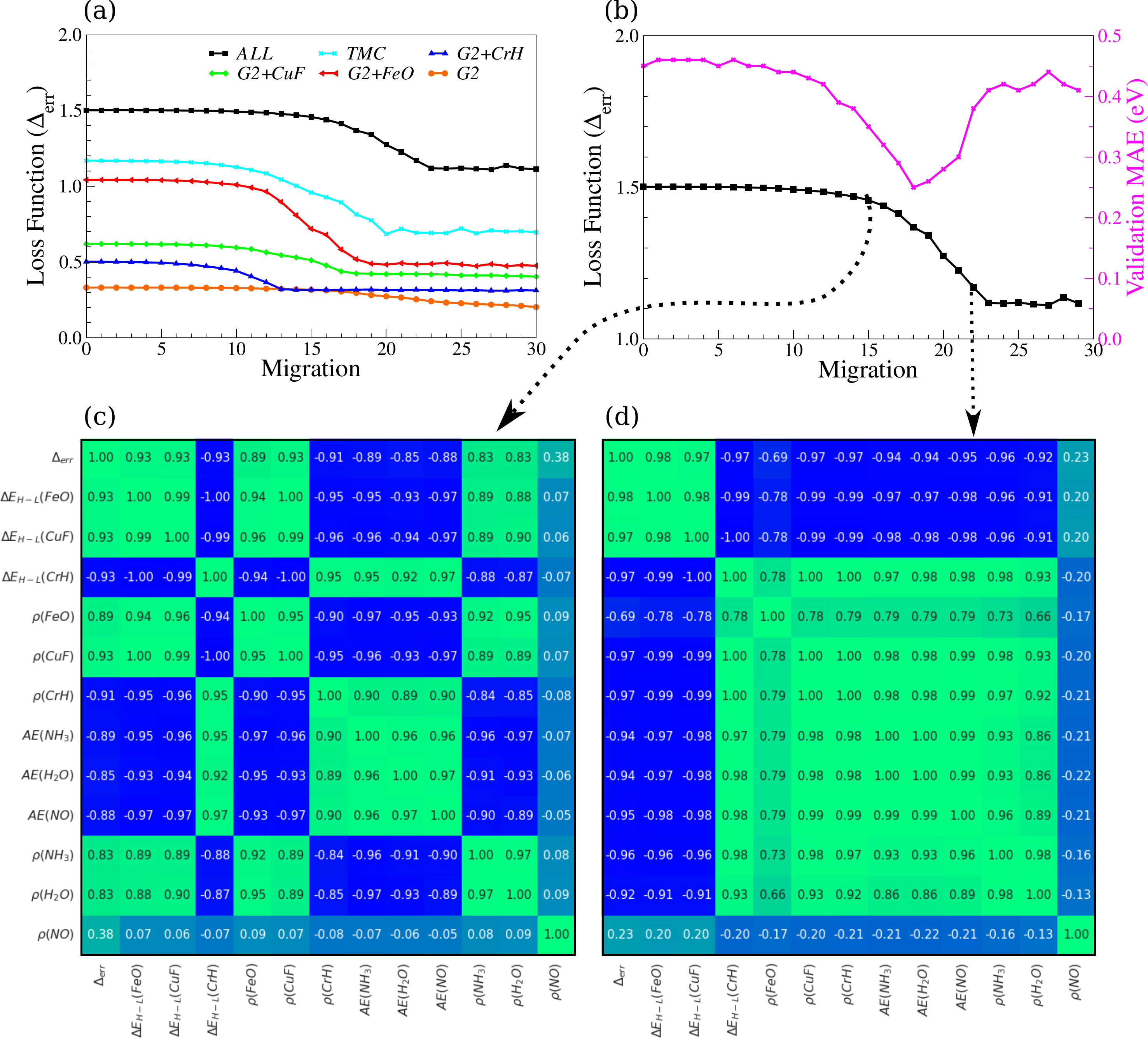}
    \caption{(a) Evolution of the loss function values during the training process, for the six different training sets considered here. (b) Comparison between the evolution of the loss function and that of the performance in the validation test along the migrations in the training with the complete \textit{ALL} training set. (c) and (d) show Spearman's rank correlation coefficients between the total loss function and each one of the $L^1$ distances that compose it. 
    The two correspond, respectively, to the samples visited along migrations \#\num{15} and \#\num{22}. The correlation matrix has converted in a heatmap.}
    \label{fig:composed_training}
\end{figure*}

In Figure{~}\ref{fig:composed_training}(a), the evolution of the loss function, as defined in eq.~\ref{eq:lossfunction} is drawn, where only the best solution (particle) at each migration is reported. After $30$ migrations, the solution is converged for every run except for \textit{G2}. For this training set, a monotonic decrease of the loss function is observed until $36$ migrations. A different behavior is observed for the training sets including \textit{TMC}, where the slow monotonic decrease is followed by a sharp decrease, and then by a stationary region.
This change in behavior is associated with solutions that cannot be  improved  \textit{via} migrations.

The training for the \textit{G2} follows path of error reduction of the atomization energy at the expense of a small increase in the error of the density (see Figures S2-S3). The loss function stops decreasing after migration \num{36} and this is associated to a limit in the  error of the atomization energy for \ce{NO} and \ce{H2O}.
When the transition metal complexes are included in the training, a fast optimization of the adiabatic energy difference is observed simultaneously with the corresponding densities, highlighting the role of accurate electronic densities in the evaluation of adiabatic energy differences\cite{song2018benchmarks,mariano2021improved}. This is seen 
 in the Spearman correlation matrices computed during the optimization for the first 2000 functionals visited for the \textit{G2+FeO} training (see Figure S6) and for the \textit{ALL} training for FeO and CuF (see Figure S15).
  In these training runs (when one or more transition metal complexes are included), the stationary region is associated with one of the $\Delta E_{H-L}$ reaching limit accuracy (\textit{i.e.}, the error associated to this quantity reaches zero). After this point, the process searches ways to optimize the other degrees of freedom (densities or atomization energies) and this leads to an increased dispersion on the error values associated with the $\Delta E_{H-L}$.

Figures{~}\ref{fig:composed_training}(c) and (d) report the heatmap for the Spearman correlation matrix between the partial errors that compose the loss function for particles sampled at migration \#$15$, \textit{i.e.} during the optimization, and at migration \#$22$, where the optimization stops, for the \textit{ALL} training. 
At migration \#$15$, in Figure{~}\ref{fig:composed_training}(c) we see that the errors on energies and densities of \ce{FeO} and \ce{CuF}, as well as the densities on the G2 molecules, are optimized simultaneously. The loss function part associated with $\Delta E_{H-L}^{\ce{CrH}}$ and $\rho_{\ce{CrH}}$ is negatively correlated with those values meaning that their error increase while the total loss function decreases. 



In the case of the \textit{ALL} and \textit{TMC} training sets, $\Delta E_{H-L}^{\ce{CuF}}$ was the first to reach a converged value close to zero. At this point of the optimization, the particles dispersion was enough to allow to find a new optimizing path for $\Delta E_{H-L}^{\ce{FeO}}$, as seen in the top left green block in Figure{~}\ref{fig:composed_training}(d) (positive correlation between these quantities and the total loss function). 
The second block in Figure{~}\ref{fig:composed_training}(d) showing strong positive correlation is negatively correlated with $\Delta_\textup{err}$.

\subsection{Validation}
\subsubsection{The performance on \ce{Fe} complexes}

 A validation test is performed by computing the errors (MEA) with respect to the CASPT2/CC adiabatic energy difference reported by Mariano et al.\cite{mariano2021improved} This is shown in the magenta curve of Figure{~}\ref{fig:composed_training}(b) where each point correspond to the best solution for each migration obtained using the \textit{ALL} training (black curve of the same Figure). The error on the validation reaches a minimum at step \#\num{18} while the loss function still decreases. The functional that performs the best on the validation test is from now on called $ANN^{ALL}_{val}$, while the one that best performs on the loss function is $ANN^{ALL}_{lf}$. Table{~}\ref{tab:validation} gathers the $\Delta E$ computed using the seven ANN-functionals (from the six training sets along with the validation). The performance of other DFT functionals, together with two recent ANN-based functionals, the one from ref. \cite{nagai2022machine}  named hereafter NAS from Nagai--Akashi--Sugino, and the DM21 one\cite{kirkpatrick2021pushing}, are also reported for comparison in table{~}\ref{tab:validation}.

\begin{table*}[htb]
\centering
\caption{Performance of the seven ANN-based exchange and correlation functionals in the prediction of the adiabatic energy differences of the validation set. HS corresponds to $S=2$ and LS to $S=0$. Other local, semilocal, hybrid and recent ANN-based functionals are also reported for comparison. The reference values are from CASPT2/CC calculations and are taken from ref.~\cite{mariano2021improved}, as discussed in the text.}
\label{tab:validation}
\renewcommand{\arraystretch}{1.2}
\resizebox{\textwidth}{!}{
\begin{tabular}{rccccccccccccccc}
\multicolumn{1}{l}{ }   & \multicolumn{15}{c}{$\Delta E_{\text{H-L}}$ (eV)}    \\ \cline{2-16} 
\multicolumn{1}{l}{ }   & \rotatebox{90}{\textbf{PBE}} & \rotatebox{90}{\textbf{SCAN}} & \rotatebox{90}{\textbf{NAS$^a$}} & \rotatebox{90}{\textbf{B3LYP}} & \rotatebox{90}{\textbf{DM21}} & \rotatebox{90}{\textbf{R2SCAN}} & \multicolumn{1}{c}{\rotatebox{90}{\textbf{PBE{[}U{]}$^b$}}} & \rotatebox{90}{\textbf{$ANN^{ALL}_{lf}$}} & \rotatebox{90}{\textbf{$ANN^{ALL}_{val}$}} & \rotatebox{90}{\textbf{$ANN^{G2+CrH}$}} & \rotatebox{90}{\textbf{$ANN^{G2+CuF}$}} & \rotatebox{90}{\textbf{$ANN^{G2+FeO}$}} & \rotatebox{90}{\textbf{$ANN^{G2}$}}  & \multicolumn{1}{c}{\rotatebox{90}{\textbf{$ANN^{TMC}$}}}  & \rotatebox{90}{\textbf{REF.$^c$}} \\ \hline
{\ce{[Fe(H2O)6]^2+}} & -1.17 & -0.81 & -0.74 & -1.44 & -0.63 & -1.36 & -1.50 & -2.51 & -1.83  & -1.03 & -2.52 & -3.06 & 3.69 & -1.36 & -1.83 \\ %
{\ce{[Fe(NH3)6]^2+}} & -0.06 &  0.21 &  0.21 & -0.58 &  0.26 & -0.29 & -0.44 & -1.27 & -0.71  & -0.01 & -1.27 & -1.73 & 3.75 & -0.29 & -0.64 \\ %
{\ce{[Fe(NCH)6]^2+}} &  1.14 &  0.89 &  0.90 & -0.21 &  0.64 &  0.41 &  0.21 & -0.39 &  0.08  &  0.64 & -0.41 & -0.79 & 3.77 &  0.41 & -0.16 \\ %
{\ce{[Fe(PH3)6]^2+}} &  2.69 &  2.09 &  2.21 &  0.62 &  2.24 &  1.91 &  1.81 &  2.05 &  1.94  &  1.89 &  2.08 &  2.16 & 0.78 &  1.90 &  2.54 \\ %
{\ce{[Fe(CO)6]^2+}}  &  3.63 &  2.86 &  2.93 &  1.25 &  2.35 &  2.58 &  2.64 &  2.42 &  2.50  &  2.60 &  2.43 &  2.36 & 3.05 &  2.57 &  2.02 \\ %
{\ce{[Fe(NCH)6]^2+}} &  4.11 &  3.38 &  3.42 &  1.86 &  2.99 &  3.06 &  3.23 &  2.93 &  2.99  &  3.10 &  2.95 &  2.87 & 3.24 &  3.05 &  2.87 \\ \hline
\multicolumn{1}{c}{\textbf{MAE$^c$}} & \textit{0.92} & \textit{0.79} & \textit{0.80}& \textit{0.70} & \textit{0.61} & \textit{0.46}  & \multicolumn{1}{c}{\textit{0.44}}    & \textit{0.41} & \textit{0.25}  & \textit{0.62}  & \textit{0.42}  & \textit{0.61}  & \textit{2.83} & \multicolumn{1}{c}{\textit{0.46}} & \multicolumn{1}{l}{ }\\ \cline{1-15}
\multicolumn{16}{l}{\footnotesize $^a$ NAS is the functional obtained by R. Nagai, R. Akashi, and O. Sugino.\cite{nagai2022machine}}  \\ 
\multicolumn{16}{l}{\footnotesize $^b$ PBE{[}U{]} represents the use of Hubbard $U_{sc}$-corrected density in the PBE functional.\cite{mariano2021improved}}  \\ 
\multicolumn{16}{l}{\footnotesize $^c$ Reference values come from CASPT2/CC.\cite{mariano2021improved}}  \\
\end{tabular}
}
\end{table*}

The performance of PBE, PBE[U], SCAN, and B3LYP was already discussed in previous studies\cite{song2018benchmarks,mariano2021improved}. Here, consistent with previous results,  MAE of $0.92$, $0.79$ and \SI{0.70}{\electronvolt} are found. The values computed using the Hubbard-U density corrected scheme using linear-response values of $U$, named PBE[U], are also reported for comparison. These are taken from ref.\cite{mariano2021improved} where the method was shown to yield the smallest error among $19$ different density functional approaches. The PBE[U] associated MAE is \SI{0.44}{\electronvolt}, and larger errors are observed for TPSSh (MAE$=$\SI{0.49}{\electronvolt}), PBE0 (MAE$=$\SI{0.64}{\electronvolt}) and \mbox{M06-2X} (MAE$=$\SI{2.41}{\electronvolt}).

 It is worth noting here that R2SCAN (MAE$=$\SI{0.49}{\electronvolt}) yields a significant improvement over SCAN (MAE$=$\SI{0.79}{\electronvolt}) in terms of $\Delta E$, with a performance close to PBE[U]. This can be related to the improved numerical stability of the former (see section S1.1 in the SI). The two ANN-based functionals developed recently\cite{nagai2022machine,kirkpatrick2021pushing} exhibit a performance worse than R2SCAN: the ANN-based DM21 using range separated hybrids yields MAE=\SI{0.61}{\electronvolt}; the NAS\cite{nagai2022machine} functional yields  MAE$=$\SI{0.80}{\electronvolt} which is similar to SCAN, consistent with the fact that it is built as a correction over SCAN to improve the performance of atomization energies of light molecules, like the ones in the G2/G3 dataset. It is important at this point to note that $ANN^{G2}$, trained here only on the \textit{G2} set (\ce{H2O}, \ce{NH3}, \ce{NO}), is the worst performer with a MAE of \SI{2.83}{\electronvolt}. 
This shows a far worse performance than NAS which is trained on similar molecules, i.e.~\ce{H2O}, \ce{NH3}, \ce{CH2}. For atomization energies on the G2 dataset, NAS gives MAE=\SI{0.156}{\electronvolt}\cite{nagai2020completing}, while for SCAN MAE$=$\SI{0.269}{\electronvolt}; for $ANN^{G2}$ MAE=\SI{0.072}{\electronvolt}, while for R2SCAN MAE$=$\SI{0.127}{\electronvolt} (see Figure SX). The improved description of the atomization energies for $ANN^{G2}$ at the expense of the performance on the adiabatic energy differences points to a possible overfitting.

This does not occur in the case of $ANN^{ALL}$: this functional yields an improved prediction of the adiabatic energy difference (MAE$=$\SI{0.41}{\electronvolt}) with only a marginal worsening for the atomization energies (\SI{0.15}{\electronvolt}, see Table S1).

Among all the developed functionals, \textit{ALL} and \textit{G2+CuF}
 yield the best results with an improvement over R2SCAN, with the former, \textit{i.e.}~being the best performer. In Table{~}\ref{tab:validation}, the performance of $ANN^{ALL}_{lf}$ and $ANN^{ALL}_{val}$is displayed.
While the validation set only includes octahedral Fe complexes involving a $S=0$ to $S=2$ spin state change, $ANN^{ALL}_{lf}$ should aim at a lager diversity in terms of transition metal cation, spin state, and electronic structure. $ANN^{ALL}_{val}$, with its low MAE of \SI{0.25}{\electronvolt}, showcases how our results can be biased towards a specific situation.

\subsubsection{Generalization on other metallic complexes}
Another set of molecules is now used with different types of transition metal atoms and varying local coordinations in order to evaluate the performance of $ANN^{ALL}_{lf}$ as compared to the biased $ANN^{ALL}_{val}$. 
Nine spin state splitting values from six molecular complexes are considered from the work of Pierloot et al.~\cite{pierloot2017spin,phung2018toward}. This set contains Fe complexes, such as \ce{FeL2} and \ce{FeL2SH}, in a square-planar and square-pyramidal configuration respectively, \ce{MnL2}, the octahedral complex \ce{[Co(NCH)6]^{2+}},
and the two organometallic complexes, \ce{MnCp_2}, and \ce{NiCp(acac)} (L2 being the bidendate \ce{C3N2H_5^-}, Cp$=$cyclopentadienyl, acac$=$acetylacetonate).

The predictions for adiabatic energy differences obtained using $ANN^{ALL}_{lf}$ and $ANN^{ALL}_{val}$ is reported in  Table{~}\ref{tab:generality} together with other known functionals such as TPSS, R2SCAN, PBE, PBE[U] and DM21. Coupled cluster-corrected CASPT2 (CASPTE/CC) values from Pierloot \textit{et al.} are taken as here reference\cite{pierloot2017spin}.

\begin{table*}[htb]
\caption{Performance of different exchange and correlation functionals for the prediction of adiabatic energy differences of the test set. Some spin configurations could not be converged depending on spin state and thus a few values from the table are missing, and marked as ``-''. }
\label{tab:generality}
\renewcommand{\arraystretch}{1.2}
\begin{tabular}{rccccccccc}
                         &      & \multicolumn{8}{c}{$\Delta E_{\text{H-L}}$ (eV)}  \\ \cline{3-10} 
                         &  2S+1 & $ANN^{ALL}_{lf}$   & $ANN^{ALL}_{val}$  & R2SCAN  & TPSS   & PBE    & PBE{[}U{]}     & DM21 & REF.$^b$  \\ \hline
FeL$_2$                  & $1\rightarrow5$  & -1.439  & -0.944  & -0.611  & 0.227  & -     & -1.096  &  -1.968  & -1.487 \\ 
                         & $3\rightarrow5$  & 0.193   & 0.513   & 0.746   & 1.259  & 1.245  & 0.756  &  0.814   & 0.213 \\ 
MnL$_2$                  & $2\rightarrow6$  & -2.078  & -1.427  & -0.957 & -0.055  & -0.111  & -0.961 &  -    & -1.782  \\
                         & $4\rightarrow6$  & -0.061  & 0.207  & 0.414  & 0.835  & 0.919  & 0.491  &  -0.739   & -0.455 \\
FeL$_2$SH                & $2\rightarrow6$  & -0.464  & 0.081  & 0.470  & -0.416  & 1.087  & -      &   -      & 0.399 \\ 
                         & $4\rightarrow6$  & -0.488  & -0.130  & 0.123  & 0.588  & 0.521  & 0.245  &   -       & -0.017 \\
{[}Co(NCH)$_6${]}$^{2+}$ & $2\rightarrow4$  & -0.372  & -0.170  & -0.027  & 0.31   & 0.476  & 0.165  &  0.057   & -0.581 \\ 
NiCp(acac)               & $1\rightarrow3$  & -0.204  & -0.177  & -0.157  & 0.046  & 0.163  & 0.095  &  0.414   & 0.117 \\  
MnCp$_2$                 & $2\rightarrow6$  & -0.220  & -0.354  & 0.184  & 0.649  & 0.981  & 0.432  &   -       & 0.304 \\ \hline 
\textit{MAE$^b$} & \textit{} & \textit{0.349} & \textit{0.406} & \textit{0.474} & \textit{0.945} & \textit{0.885$^a$}& \textit{0.482$^a$} & \textit{0.460$^a$} & \multicolumn{1}{l}{}\\ \cline{1-9}
\multicolumn{10}{l}{\footnotesize $^a$ Average computed over available values.}  \\
\multicolumn{10}{l}{\footnotesize $^b$ Reference values come from CASPT2/CC.\cite{pierloot2017spin}}  \\
\end{tabular}
\end{table*}

As shown above for the validation set, R2SCAN (MAE=0.474 eV) substantially improves over PBE (MAE=0.885 eV) and TPSS (MAE=0.945 eV). The best performers, $ANN^{ALL}_{lf}$ and $ANN^{ALL}_{val}$, further improve over R2SCAN with MAE of 0.349 eV and 0.406 eV, respectively. We note a better performance of DM21 (MAE=0.460 eV) for this set of molecules as compared with the octahedral complexes presented above. Yet, we note here that owing to the heavy architecture of the DM21 functional (it is defined pointwise using Hartree-Fock, range-separated hybrid and LDA energy densities with weights computed as ANN with hidden layers containing up to $256$ neurons), the calculation of adiabatic energy differences is costly and the SCF is sometimes difficult to converge. This is the case here where the SCF could not converge for MnL$_2$ ($2S+1=4 \textup{ and } 6$), FeL$_2$SH ($2S+1=2 \textup{ and } 4$), and MnCp$_2$ ($2S+1=6$) despite starting the calculation using an already converged density (R2SCAN here) as suggested in the DM21 documentation\cite{DM21git}.

Among the studied molecules, the square planar compounds \ce{MnL2} and \ce{FeL2} result in the largest error for R2SCAN of up to \SI{0.9}{\electronvolt}, a performance that is significantly improved by both the functionals trained with $ALL$. 
Also, a good performance is seen for $ANN^{ALL}_{lf}$ to describe complexes far from the training set in terms of geometry and choice of atoms, such as \ce{[Co(NCH)_6]^{2+}} (error of \SI{0.21}{\electronvolt}). 

The curves shown in Figure{~}\ref{fig:composed_training}(b) may convey the message of an over-fitting during the \textit{ALL} training, as seen from the increase of the validation error. During the optimization on the training set, the performance for the chosen validation set gets worse after migration \#18.
Yet, the $ANN^{ALL}_{lf}$ functional outperforms $ANN^{ALL}_{val}$ in the generality test (see Table~\ref{tab:generality}). The fact that the best fitted functional for the training set best performs also for the generality test implies a good transferability of the functional. This can be associated also to the physical limits imposed during the training.


\section{Conclusion}

The ANN-based exchange and correlation functional in DFT developed in this work outperforms most known local, semilocal, and meta-GGA functionals in the prediction of adiabatic energy differences for transition metal complexes. 
It is built as a physically-constrained meta-GGA artificial neural-network correction to R2SCAN and trained over a small dataset of electronic densities, atomization energies and/or adiabatic energy differences. 
Our results show a good performance of the non gradient-based bioinspired training algorithm PSO where the optimization can reach a hard limit, as in the case of the adiabatic energy differences of \ce{CuF}, \ce{FeO}, and \ce{CrH}.
Six functionals were obtained by using different subsets of the training molecules. The functional optimized by using the whole training set, $ANN^{ALL}_{lf}$ (MAE$=$\SI{0.41}{\electronvolt}), and that obtained using the $G2+CuF$, $ANN^{G2+CrF}$ (MAE$=$\SI{0.42}{\electronvolt}), yield a performance on the validation test superior to the best performer R2SCAN (MAE$=$\SI{0.44}{\electronvolt}). Such a validation test performed on octahedral iron complexes shows that our functional substantially improves over recent ANN-based functionals, such as the meta-GGA NAS and the range-separated hybrid DM21.
The best performer on the validation test, $ANN^{ALL}_{lf}$, best performs also on the generality test with a MAE$=$\SI{0.349}{\electronvolt}.
Finally, we show that by changing the the criteria used to select the functional (from the loss function itself to a validation metric), we were able to bias the results to improve the performance on a specific set of data. For example, the best performer on the validation test, named $ANN^{ALL}_{val}$, yields a MAE as small as \SI{0.25}{\electronvolt}, but the associated MAE on the generality test is larger (MAE$=$\SI{0.41}{\electronvolt}) than for the $ANN^{ALL}_{lf}$. 
Overall, the results of this study highlight the impact of machine learning approaches to enhance the accuracy of approximate density functional theory through the development of optimized functionals.

\begin{acknowledgement}
Computational resources were provided by the CINES and IDRIS under Project No. INP2227/72914, as well as CIMENT/GRICAD for computational resources. This work was performed within the framework of the Centre of Excellence of Multifunctional Architectured Materials CEMAM-ANR-10-LABX-44-01 funded by the ``Investments for the Future'' Program. This work has been partially supported by MIAI@Grenoble Alpes (ANR-19-P3IA-0003). The authors thank Lucia Reining for fruitful discussions. Discussions within the French collaborative network in artificial intelligence in materials science GDR CNRS 2123 (IAMAT) are also acknowledged.
\end{acknowledgement}

\begin{suppinfo}
Experimental procedures and characterization data for all new compounds. The class will automatically add a sentence pointing to the information on-line:
\end{suppinfo}

\bibliography{referencias}

\end{document}


\tableofcontents

\section{Preliminary tests}
\subsection{Training over SCAN using Water and Iron Oxide}

A set of optimization runs was performed to study the role of the density in the optimization of the adiabatic energy difference. To do so, the same methodology employed on the main training was used here, with the exception of two features. The first difference is that the $\Delta_{\text err}$ function in the Eq.~$7$ was modified to include either (1) the error on the energy (i.e.~the adiabatic energy difference for \ce{FeO} and the atomization energy for \ce{H2O}), (2) the error on the density error only, and (3) the sum of both terms. The second difference is that here SCAN is used as the reference functional as in Nagai-Akashi-Sugino (NAS) functional.\cite{nagai2022machine} Figure{~}\ref{fig:scan_test} shows the evolution of the loss function computed as error on the energy or the electronic density for the three optimization runs mentioned above (1), (2), and (3). 

\begin{figure}[H]
    \centering
    \includegraphics[width=0.95\linewidth]{figures_si/preliminarytraining.png}
    \caption{Training runs performed over SCAN for FeO (upper panels) and H$_2$O (lower panels) using three different definitions of the loss function: (1) the adiabatic energy difference (orange dots), (2) the electronic desity (yellow dots), and (3) both (dark blue dots) ($\Delta_{\text err}$).}
    \label{fig:scan_test}
\end{figure}

The results show a good correlation between the error in the adiabatic energy difference and the error on the electronic density for \ce{FeO}. 
When the training is performed using only the density, a quick optimization evolves to a region with an associated decrease of the error on the energy. However, the SCF convergence was unstable and the optimization could not be pursued after $3000$ cycles.

Interestingly, for \ce{FeO}, small changes in the parameters of the functional can yield very different energies as seen in Figure{~}\ref{fig:scan_test}. This is possibly related to the use of SCAN as discussed in the main paper. Such behavior is not found when using R2SCAN as expected from the smooth dependence of $\epsilon_{XC}^{R2SCAN}$ on $\tau$\cite{furness2020accurate}. 

For \ce{H2O}, there is no correlation between the error in the energy and in the density. Since the errors in atomization energy and in the density are independent, an optimization path can show improvement in one quantity while getting worse in the other, as found for the training using the atomization energy alone and the $\rho$ alone. When both quantities are used for the optimization, the performance for the atomization energy improves (although the error in the density increases). This results confirms the general outcome of this work: the energetic properties dominate during the training process, with the density acting as an auxiliary quantity that eventually allow for a further reduction of the error.

\section{Correlation Analysis}
\subsection{G2 training}

\begin{figure}[H]
    \centering
    \includegraphics[width=0.6\linewidth]{figures_si/g2_pca.png}
    \hspace{0.05\linewidth}
    \raisebox{0.5\height}{\includegraphics[width=0.3\linewidth]{figures_si/g2_correl.png}}
    \caption{Principal Component Analysis of the different terms of the loss function along the training with \textit{G2} (left Figure). Spearman correlation matrix between these terms and whole the loss functions, $\Delta_{\text err}$ (right Figure).}
    \label{fig:g2_pca}
\end{figure}

\begin{figure}[H]
    \centering
    \includegraphics[height=0.2\linewidth]{figures_si/g2_correl_init.png}
    \hspace{0.01\linewidth}
    \includegraphics[height=0.2\linewidth]{figures_si/g2_correl_fin.png}
    \caption{Spearman correlation matrices between the whole loss function and its individual terms while training with \textit{G2} computed during the first $2000$ (left Figure) and the last $2000$ iterations (functionals) (right Figure), highlighting the differences between the initial and the final behaviors.}
    \label{fig:g2_correl}
\end{figure}

\begin{figure}[H]
    \centering
    \includegraphics[width=0.8\linewidth]{figures_si/g2_pairplot.png}
    \caption{Pairplot showing the relationship between the different terms of the loss function during the optimization for the \textit{G2}. The Index axis orders the samples obtained during the training. In a pairplot, the x axis is shared among plots lying in the same column and the y axis is shared among plots on the same line, with the exception of the main diagonal where the histogram of each quantity is shown.}
    \label{fig:g2_pairplot}
\end{figure}8

\subsection{G2+FeO training}

\begin{figure}[H]
    \centering
    \includegraphics[width=0.5\linewidth]{figures_si/feo_pca.png}
    \hspace{0.05\linewidth}
    \raisebox{0.1\height}{\includegraphics[width=0.4\linewidth]{figures_si/feo_correl.png}}
    \caption{Principal Component Analysis of the different terms of the loss function along the training with \textit{G2+FeO} (left Figure). Spearman correlation matrix between these terms and whole the loss functions, $\Delta_{\text err}$ (right Figure).}
    \label{fig:feo_pca}
\end{figure}

\begin{figure}[H]
    \centering
    \includegraphics[height=0.4\linewidth]{figures_si/feo_correl_init.png}
    \hspace{0.01\linewidth}
    \includegraphics[height=0.4\linewidth]{figures_si/feo_correl_fin.png}
    \caption{Spearman correlation matrices between the whole loss function and its individual terms while training with \textit{G2+FeO} computed during the first $2000$ (left Figure) and the last $2000$ iterations (functionals) (right Figure), highlighting the differences between the initial and the final behaviors.}
    \label{fig:feo_correl}
\end{figure}

\begin{figure}[H]
    \centering
    \includegraphics[width=0.9\linewidth]{figures_si/feo_pairplot.png}
    \caption{Pairplot showing the relationship between the different terms of the loss function during the optimization for the \textit{G2+FeO}. The Index axis orders the samples obtained during the training. In a pairplot, the x axis is shared among plots lying in the same column and the y axis is shared among plots on the same line, with the exception of the main diagonal where the histogram of each quantity is shown.}
    \label{fig:feo_pairplot}
\end{figure}

\subsection{G2+CrH training}

\begin{figure}[H]
    \centering
    \includegraphics[width=0.5\linewidth]{figures_si/crh_pca.png}
    \hspace{0.05\linewidth}
    \raisebox{0.1\height}{\includegraphics[width=0.4\linewidth]{figures_si/crh_correl.png}}
    \caption{Principal Component Analysis of the different terms of the loss function along the training with \textit{G2+CrH} (left Figure). Spearman correlation matrix between these terms and whole the loss functions, $\Delta_{\text err}$ (right Figure).}
    \label{fig:crh_pca}
\end{figure}

\begin{figure}[H]
    \centering
    \includegraphics[height=0.4\linewidth]{figures_si/crh_correl_init.png}
    \hspace{0.01\linewidth}
    \includegraphics[height=0.4\linewidth]{figures_si/crh_correl_fin.png}
    \caption{Spearman correlation matrices between the whole loss function and its individual terms while training with \textit{G2+CrH} computed during the first $2000$ (left Figure) and the last $2000$ iterations (functionals) (right Figure), highlighting the differences  between the initial and the final behaviors.}
    \label{fig:crh_correl}
\end{figure}

\begin{figure}[H]
    \centering
    \includegraphics[width=0.9\linewidth]{figures_si/crh_pairplot.png}
    \caption{Pairplot showing the relationship between the different terms of the loss function during the optimization for the \textit{G2+CrH}. The Index axis orders the samples obtained during the training. In a pairplot, the x axis is shared among plots lying in the same column and the y axis is shared among plots on the same line, with the exception of the main diagonal where the histogram of each quantity is shown.}
    \label{fig:crh_pairplot}
\end{figure}

\subsection{G2+CuF training}

\begin{figure}[H]
    \centering
    \includegraphics[width=0.5\linewidth]{figures_si/cuf_pca.png}
    \hspace{0.05\linewidth}
    \raisebox{0.1\height}{\includegraphics[width=0.4\linewidth]{figures_si/cuf_correl.png}}
    \caption{Principal Component Analysis of the different terms of the loss function along the training with \textit{G2+CuF} (left Figure). Spearman correlation matrix between these terms and whole the loss functions, $\Delta_{\text err}$ (right Figure).}
    \label{fig:cuf_pca}
\end{figure}

\begin{figure}[H]
    \centering
    \includegraphics[height=0.4\linewidth]{figures_si/cuf_correl_init.png}
    \hspace{0.01\linewidth}
    \includegraphics[height=0.4\linewidth]{figures_si/cuf_correl_fin.png}
    \caption{Spearman correlation matrices between the whole loss function and its individual terms while training with \textit{G2+CuF} computed during the first $2000$ (left Figure) and the last $2000$ iterations (functionals) (right Figure), highlighting the differences  between the initial and the final behaviors.}
    \label{fig:cuf_correl}
\end{figure}

\begin{figure}[H]
    \centering
    \includegraphics[width=0.9\linewidth]{figures_si/cuf_pairplot.png}
    \caption{Pairplot showing the relationship between the different terms of the loss function during the optimization for the \textit{G2+CuF}. The Index axis orders the samples obtained during the training. In a pairplot, the x axis is shared among plots lying in the same column and the y axis is shared among plots on the same line, with the exception of the main diagonal where the histogram of each quantity is shown.}
    \label{fig:cuf_pairplot}
\end{figure}

\subsection{ALL training}

\begin{figure}[H]
    \centering
    \raisebox{0.05\height}{\includegraphics[width=0.45\linewidth]{figures_si/all_pca.png}}
    \hspace{0.05\linewidth}
    \includegraphics[width=0.45\linewidth]{figures_si/all_correl.png}
    \caption{Principal Component Analysis of the different terms of the loss function along the training with \textit{ALL} (left Figure). Spearman correlation matrix between these terms and whole the loss functions, $\Delta_{\text err}$ (right Figure).}
    \label{fig:all_pca}
\end{figure}

\begin{figure}[H]
    \centering
    \includegraphics[height=0.5\linewidth]{figures_si/all_correl_init.png}
    \hspace{0.01\linewidth}
    \includegraphics[height=0.5\linewidth]{figures_si/all_correl_fin.png}
    \caption{Spearman correlation matrices between the whole loss function and its individual terms while training with \textit{ALL} computed during the first $2000$ (left Figure) and the last $2000$ iterations (functionals) (right Figure), highlighting the differences between the initial and the final behaviors.}
    \label{fig:all_correl}
\end{figure}

\begin{figure}[H]
    \centering
    \includegraphics[width=0.9\linewidth]{figures_si/all_pairplot.png}
    \caption{Pairplot showing the relationship between the different terms of the loss function during the optimization for the \textit{ALL}. The Index axis orders the samples obtained during the training. In a pairplot, the x axis is shared among plots lying in the same column and the y axis is shared among plots on the same line, with the exception of the main diagonal where the histogram of each quantity is shown.}
    \label{fig:all_pairplot}
\end{figure}

\subsection{TMC training - only diatomic molecules with transition metals}

For completeness, a training was done including only the molecules of the \textit{ALL} training set that posses a transition metal. As the training with \textit{ALL} was dominated by the optimizations on the values of $\Delta E_{H-L}$, we confirmed that this extra training has a performance very similar to the one with the complete set, but with a slightly better validation performance of $ANN^{ALL}_{lf}$ ($MAE=0.41$) in relation to $ANN^{MC}$ ($MAE=0.46$). In other hand, even if we see the same absolute error and the training evolution is similar, its important to highlight that the functionals we obtained are very different, with $ANN^{ALL}_{lf}$ tending to underestimate $\Delta E_{H-L}$ ($ME=-0.26$) on the validation set, while $ANN^{MC}$ overestimate almost all cases ($ME=0.25$, only underestimating $\Delta E_{H-L}$ for \ce{[Fe(PH3)6]^{2+}}.

\begin{figure}[H]
    \centering
    \raisebox{0.05\height}{\includegraphics[width=0.45\linewidth]{figures_si/mc_pca.png}}
    \hspace{0.05\linewidth}
    \includegraphics[width=0.45\linewidth]{figures_si/mc_correl.png}
    \caption{On the left, Principal Component Analysis of the many different possible error components on the loss function along the training with \textit{MC}. In the right, the Spearman correlation matrix between those components and the loss functions ($\Delta_{\text err}$). Lines and columns set with high brightness are associated with variables not considered on the loss function, but computed for easier comparison with $ALL$ (Figure{~}\ref{fig:all_pca}).}
    \label{fig:mc_pca}
\end{figure}

\begin{figure}[H]
    \centering
    \includegraphics[height=0.5\linewidth]{figures_si/mc_correl_init.png}
    \hspace{0.01\linewidth}
    \includegraphics[height=0.5\linewidth]{figures_si/mc_correl_fin.png}
    \caption{Spearman correlation matrices between the whole loss function and its individual terms while training with \textit{ALL} computed during the first $2000$ (left Figure) and the last $2000$ iterations (functionals) (right Figure), highlighting the differences on between the initial and the final behaviors. Lines and columns set with high brightness are associated with variables in the loss function set to zero during the optimization, but computed for easier comparison with $ALL$ (Figure{~}\ref{fig:all_correl}).}
    \label{fig:mc_correl}
\end{figure}

\begin{figure}[H]
    \centering
    \includegraphics[width=0.9\linewidth]{figures_si/mc_pairplot_ini.png}
    \caption{Pairplot showing the relationship between the different terms of the loss function during the first 2000 functionals visited along the \textit{TMC} training.  Here, the low correlation between $\Delta_{\textup err}$ and the $L^1$ error on $\rho$ for \ce{CrH} (see the left plot on Figure{~}\ref{fig:mc_correl}) is explained via a change of behavior of this variable, going from positive to negative correlation inside this interval.}
    \label{fig:mc_pairplot}
\end{figure}

\section{Generality test: atomization energies for the G2 and G3 datasets}

A subset of $66$ small molecules inside the G2 and G3 datasets was selected to perform a generality test of atomization energies using the ANN-based functionals. The list of computed molecules is:  \ce{H2}, \ce{LiH}, \ce{Li2}, \ce{BeH}, \ce{CH}, \ce{CH3}, \ce{CH4}, \ce{NH}, \ce{NH2}, \ce{NH3}, \ce{OH}, \ce{OH2}, \ce{C2H}, \ce{C2H2}, \ce{C2H4}, \ce{C2H6}, \ce{CN}, \ce{HCN}, \ce{FH}, \ce{LiF}, \ce{N2}, \ce{CO}, \ce{HCO}, \ce{H2CO}, \ce{NO}, \ce{O2}, \ce{H2O2}, \ce{N2O}, \ce{CO2}, \ce{F2}, \ce{NO2}, \ce{O3}, \ce{F2O}, \ce{SiH3}, \ce{SiH4}, \ce{Na2}, \ce{BF3}, \ce{PH2}, \ce{PH3}, \ce{NF3}, \ce{SiO}, \ce{HS}, \ce{SH2}, \ce{SC}, \ce{CF4}, \ce{ClH}, \ce{SO}, \ce{C2F4}, \ce{COS}, \ce{ClO}, \ce{HOCl}, \ce{AlF3}, \ce{SO2}, \ce{FCl}, \ce{Si2}, \ce{ClNO}, \ce{NaCl}, \ce{PF3}, \ce{P2}, \ce{SiF4}, \ce{ClF3}, \ce{S2}, \ce{CS2}, \ce{Cl2}, \ce{BCl3},and \ce{CCl4}. All are neutral and considered in their ground state. The result is shown in Figure{~}\ref{fig:g2_valid}.

\begin{figure}[H]
    \centering
    \includegraphics[width=0.5\linewidth]{figures_si/g2valid.eps}
    \caption{Sample predictions of Atomization Energy (AE) made by R2SCAN and some of the functionals on this work.}
    \label{fig:g2_valid}
\end{figure}

\begin{table}[H]
\caption{Errors on AE in the G2 subset.}
\begin{tabular}{rcccc}
\multicolumn{1}{l}{} & \multicolumn{4}{c}{AE (eV)}                          \\ \cline{2-5} 
\multicolumn{1}{l}{} & R2SCAN & $ANN^{ALL}_{val}$ & $ANN^{ALL}_{lf}$ & $ANN^{G2}$ \\ \hline
ME      &   0.12&	0.13	&0.15&	-0.03 \\
MAE     &   0.13&	0.14	&0.15&	0.07 \\
SD      &   0.11&	0.12	&0.13&	0.09 \\ \hline
\end{tabular}
\end{table}

The results from $ANN^{ALL}_{lf}$ and $ANN^{ALL}_{val}$ are similar to R2SCAN reflecting the fact that the optimization enforced an improved description of the molecules with transition metal atoms. The mean percentage error is $31.7\%$ for R2SCAN, $34.0\%$ for $ANN^{ALL}_{val}$, and $37.9\%$ for $ANN^{ALL}_{lf}$.

A different result is found for the $ANN^{G2}$ functional: trained on densities and atomization energies of light molecules, this functional greatly improves over R2SCAN registering a mean percentage error of $-4.0\%$. The mean absolute percentile error (MAPE) of R2SCAN ($MAPE=35.1\%$) is also reduced significantly in $ANN^{G2}$ ($MAPE=20.2\%$).


\section{Optimized densities: the transition metal molecules from training set}

We employ the $L^1$ distance metric, also used in this work as part of the loss function for the density (4$^{th}$ right-hand side term of eq.~7 in the main paper), to analyse the performance of the optimized functional as compared to other known ones, in the prediction of the electronic density. This metric is computed by comparing sum of the integrated difference in LS and HS (see text). The distance matrix heatmaps are shown in Figure{~}\ref{fig:dens_feo}, \ref{fig:dens_cuf}, and \ref{fig:dens_crh}.



\begin{figure}[H]
    \centering
    \includegraphics[width=\linewidth]{figures_si/dist_dens_feo.pdf}
    \caption{$L^1$ distances between densities computed using different density functional approaches and the CCSD(T) OptOrb densities for reference, summed over high and low spin state for \ce{FeO}. left panel: heatmap of the distance matrix reported in such a way that  cluster structure is emphasized. Right panel: 2D projection of the same matrix using Local Linear Embedding. The colour code for both figure is the same and on the left plot the distance is computed with respect to CCSD(T).}
    \label{fig:dens_feo}
\end{figure}

\begin{figure}[H]
    \centering
    \includegraphics[width=\linewidth]{figures_si/dist_dens_cuf.pdf}
    \caption{$L^1$ distances between densities computed using different density functional approaches and the CCSD(T) OptOrb densities for reference, summed over high and low spin state for \ce{CuF}. left panel: heatmap of the distance matrix reported in such a way that  cluster structure is emphasized. Right panel: 2D projection of the same matrix using Local Linear Embedding. The colour code for both figure is the same and on the left plot the distance is computed with respect to CCSD(T).}
    \label{fig:dens_cuf}
\end{figure}

\begin{figure}[H]
    \centering
    \includegraphics[width=\linewidth]{figures_si/dist_dens_crh.pdf}
    \caption{$L^1$ distances between densities computed using different density functional approaches and the CCSD(T) OptOrb densities for reference, summed over high and low spin state for \ce{CrH}. left panel: heatmap of the distance matrix reported in such a way that  cluster structure is emphasized. Right panel: 2D projection of the same matrix using Local Linear Embedding. The colour code for both figure is the same and on the left plot the distance is computed with respect to CCSD(T).}
    \label{fig:dens_crh}
\end{figure}

The results show the existence of clusters where functionals  describe the densities similarly according to the $L^1$ metric. For instance, for \ce{CrH} we see the distances matrix organized in blocks that separate hybrids from lower rung functionals. This separation is less clear for the other two molecules. Interestingly, the optimized $ANN$ functional gives a density systematically closer to CCSD(T). 
This similarity to the coupled cluster density is seen in two ways: via the small value of the $L^1$ error distance and through the similar way in which both the optimized functional and coupled cluster compare with the rest of the functionals (seen in Figure XX as  similar lines on the distances matrix and in the proximity of the points in the 2D Local Linear Embedding). This behavior suggests that  the overall density error decreases and at the same time that the density distribution is close to the CCSD(T) result. 


\section{Complete Project Flowchart}

\begin{figure*}[H]
    \centering
    \includegraphics[width=0.9\textwidth]{figures_si/flowchart.pdf}
    \caption{Schematic representation of full functional training using the approach we follow on this paper. We highlighted in blue the step inside the single point DFT calculations to compute the loss function.}
    \label{fig:abstract}
\end{figure*}

\bibliography{referencias}